%% file: root.tex
\documentclass[letterpaper, 10 pt, conference]{ieeeconf}  

\IEEEoverridecommandlockouts                              

\title{\LARGE \bf
Learning Neural Hybrid Surrogates for Gradient-Based Falsification
}

\author{Lasse Kötz and Knut Åkesson
\thanks{This work was not supported by any organization}
\thanks{Lasse Kötz and Knut Åkesson are with the Departiment of Electrical Engineering, Chalmers University of Technology, 412 96, Göteborg, Sweden {\tt\small \{kotz, knut\}@chalmers.se}}
}

\usepackage{url}
\usepackage{graphicx}
\usepackage{amsmath}
\usepackage{amssymb}
\usepackage{algorithm}
\usepackage{algpseudocode}
\usepackage{multirow}
\usepackage{booktabs}
\usepackage{adjustbox}
\usepackage{subcaption}
\usepackage{tikz}
\usepackage{xcolor}

\usetikzlibrary{positioning}
\usetikzlibrary{calc}

\newtheorem{definition}{Definition}

\begin{document}
\maketitle
\thispagestyle{empty}
\pagestyle{empty}


\input{Sections/Abstract}
\input{Sections/Introduction}
\input{Sections/Background}
\input{Sections/Method}
\input{Sections/Evaluation}

\input{Sections/Results}
\input{Sections/Conclusions}

\addtolength{\textheight}{-0cm}   




\bibliographystyle{IEEEtran}
\bibliography{root}

\end{document}

%% file: Sections/Abstract.tex
\begin{abstract}
Falsification of hybrid dynamical systems remains challenging due to mode-dependent dynamics and discrete transitions. In this work, we propose a surrogate-based falsification approach that enables hybrid systems by learning a differentiable hybrid automaton model from data. This extends previous surrogate-based falsification methods, which were limited to purely continuous dynamics. Specifically, we employ neural hybrid automata to learn both a latent mode encoder and the corresponding mode-conditioned vector fields. Once the surrogate has paired each mode with an associated vector field, the transition guards are inferred using existing trajectory data. 

The learned surrogate is subsequently subjected to a gradient-based optimal control formulation, which minimizes a smooth approximation of the safety specification to find safety violations. In the last step, an experiment with the optimal control solution is carried out on the original system to ensure soundness. 

The proposed method consistently uncovers counterexamples on a majority of evaluated benchmark specifications; on these cases, it achieves competitive or improved sample efficiency than other tools while using a reduced simulation budget.

\end{abstract}

%% file: Sections/Introduction.tex
\section{Introduction}
Recent developments in Cyber-Physical Systems (CPS) have increasingly deployed complex systems in safety-critical domains such as autonomous driving and medical devices, where failures can have severe consequences. These systems therefore require rigorous testing before deployment. 

Hybrid systems are a class of CPS characterized by a combination of continuous dynamics with discrete modes. The continuous state evolutions are conditioned on the current active mode, which in turn makes discrete changes according to a transition set and guard functions. This additional property significantly complicates the dynamics of hybrid systems, making them piecewise-continuous.

Traditional verification methods, such as formal verification and model checking, are effective for verifying the correctness of small-scale systems under test (SuT). However, they quickly become intractable as system complexity grows.
In contrast, falsification aims to find a \textit{counterexample} that violates a given safety specification - commonly expressed in signal temporal logic (STL). A counterexample consists of an input signal, along with the corresponding output signal. When a counterexample is found, the SuT can be modified. 

\begin{figure}
    \input{Figures/algorithm_schematic}
    \caption{New trajectories are acquired by experiments on the SuT using a least-robust input signal, resulting from an optimal control solution. The authenticity of the counterexample in the surrogate is always verified on the original SuT to ensure soundness.}
    \label{fig:algorithm_schematic}
\end{figure}
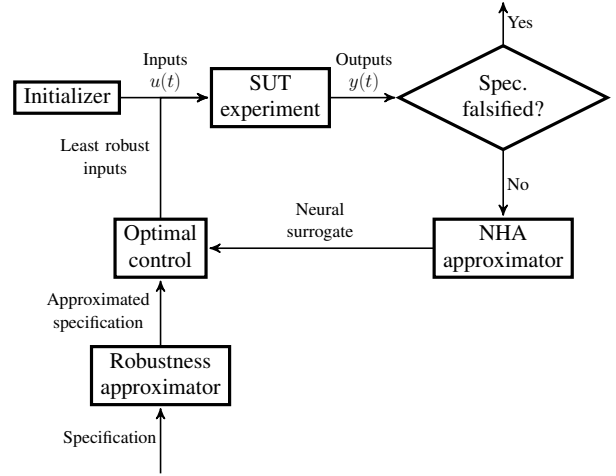

Falsification is an additional testing method that is particularly suitable when the system is treated as a black box, where internal models are either too complex for formal analysis or completely unavailable. Although falsification naturally omits many of the issues that arise in formal verification, existing methods commonly fail to leverage dynamics information, available from SuT trajectories, in their falsification strategies. 

In this paper, we build on the work in 
\cite{kotz2026optimal} to encompass the learning of hybrid systems using neural hybrid automata (NHA) and falsification by optimal control, highlighted in Figure \ref{fig:algorithm_schematic}. This is done by simultaneously training a discrete latent mode encoder network and the corresponding dynamics neural ODEs. Prior to training, trajectories are segmented into subtrajectories around switching events, using switch observers or, when unavailable, using temporal point process methods. This enables unsupervised encoder training, by minimizing reconstruction error on each segment. The trained encoder enables guard condition learning from regressing on data at transition times. Our method employs a hierarchical model selection strategy to learn guard functions. This is done by starting from a linear ansatz, only resorting to nonlinear models when the linear fit is insufficient.

The resulting surrogate NHA can be simulated cheaply compared to the original SuT and is differentiable with respect to inputs as long as the discrete mode sequence remains fixed. Candidate counterexamples are first computed on the surrogate model and subsequently verified on the original SuT. To address the inherent nonsmoothness of STL robustness semantics, we employ a smooth log-sum-exponential smoothing, yielding a differentiable robustness objective. Doing so enables a gradient-based optimal control formulation by differentiating the learned surrogate simulator and robustness value. Although hybrid automata introduce discontinuities, the surrogate remains locally differentiable in a sequence-invariant region of the input space.

We evaluate the framework on benchmarks from the ARCH-COMP 2024 falsification competition \cite{khandait2024arch} and compare against state-of-the-art tools. On the evaluated benchmarks, our method requires no more - and often fewer - SuT executions to find counterexamples. We summarize the main contributions of this paper:
\begin{itemize}
    \item We extend neural surrogate-based falsification to hybrid systems via neural hybrid automata.
    \item We show that gradient-based optimal control over the learned surrogates reduces the amount of expensive SuT executions required for falsification.
    \item We evaluate the performance on the ARCH-COMP 2024 falsification competition benchmarks and compare against other state-of-the-art methods.
\end{itemize}

%% file: Figures/algorithm_schematic.tex
\definecolor{green}{RGB}{40,188,0}
\definecolor{red}{RGB}{155,0,0}
\definecolor{darkgreen}{RGB}{8,137,0}
\definecolor{darkblue}{RGB}{9,52,226}
\definecolor{state3background}{RGB}{180, 185, 193}

\usetikzlibrary{arrows,automata,calc,positioning, shapes.geometric}

\tikzset{
    state2/.style={
           rectangle,
           draw=black, very thick,
           minimum height=2em,
           inner sep=2pt,
           minimum size=2em,
           font=\huge
           },
    state4/.style={
           diamond,
           draw=black, very thick,
           minimum height=1em,
           inner sep=1pt,
           minimum size=1em,
           font=\huge
           }
}

\begin{tikzpicture}[->,>=stealth',shorten >=1pt,auto,semithick, transform shape, scale=0.6]

\node[state2] (1) [scale=0.7]{
    \begin{tabular}{c}
        Initializer
    \end{tabular}};

\node[state2] (2) [right=2.0cm of 1, scale=0.7] {
    \begin{tabular}{c}
        SUT \\
        experiment
    \end{tabular}
    };

\node[state4] (3) [right=1.5cm of 2, scale=0.7, aspect=2] {
    \begin{tabular}{c}
        Spec. \\
        falsified?
    \end{tabular}
    };

\node[state2] (4) [below=1.5cm of 3, scale=.7] {
    \begin{tabular}{c}
         NHA  \\
         approximator
    \end{tabular}
    };

\node[state2] (6) [left=5cm of 4, scale=.7] {
    \begin{tabular}{c}
         Optimal \\ control
    \end{tabular}};

\node[state2] (7) [below=1.5cm of 6, scale=.7] {
    \begin{tabular}{c}
        Robustness \\
        approximator
    \end{tabular}
};

\coordinate[left=1.5cm of 1] (d1);
\coordinate[above=1.5cm of 3] (d2);
\coordinate[below=1.5cm of 7] (d3);
\coordinate[above=1cm of 3] (d4);
 
\path

    (1) edge [above] node[scale=.7, font=\LARGE] {
        \begin{tabular}{c}
        Inputs \\
        $u(t)$
        \end{tabular}
        } (2)
    (2) edge [above] node[scale=.7, font=\LARGE] {
        \begin{tabular}{c}
            Outputs \\
            $y(t)$
        \end{tabular}
        } (3)
    (3) edge [right] node[scale=.7, font=\LARGE] {
        No
    } 
    (4)
    (4) edge [above] node[scale=.7, font=\LARGE] {
        \begin{tabular}{c}
             Neural \\ 
             surrogate
        \end{tabular}
    } (6)

    (d3) edge [left] node[scale=.7, font=\LARGE] {
        Specification
    } (7)
    (3) edge [right] node[scale=.7, font=\LARGE] {
        Yes
    } (d4)
    (7) edge [left]
    node[scale=.7, font=\LARGE]
    {
        \begin{tabular}{c}
            Approximated  \\ specification 
        \end{tabular}
    } (6)
    ;
    
    \draw[->] (6.north) |- (2.west) node[scale=.7, font=\LARGE, left, pos=0.25] {\begin{tabular}{c}
         Least robust \\
         inputs
    \end{tabular}};

\end{tikzpicture}

%% file: Sections/Background.tex
\section{Related works}
Recent advances in falsification include optimization-based falsification methods, treating the SuT as an opaque simulator and rely on gradient-free optimization to violate a given safety specification. Methods include, simulated annealing and line-search falsification. Widely used tools such as S-TaLiRo \cite{annpureddy2011s} and Breach \cite{donze2010breach} follow this paradigm, repeatedly simulating the SuT while optimizing input parameters through stochastic search. ForeSee \cite{zhang2021effective} addresses the scale sensitivity of STL robustness by transforming the specification syntax tree and uses Monte Carlo tree search to explore the resulting structure. ATheNA \cite{formica2023search} combines manually designed and learned fitness functions, enabling domain experts to bias the exploration toward critical regions.

To guide falsification more strategically and reduce execution cost, recent techniques approximate a robustness function from data. Bayesian optimization \cite{ramezani2025falsification} learns a probabilistic mapping of the robustness as a function of the input signal parameters with an acquisition function to select new experiments. NNFal \cite{kundu2024data} trains a static depth neural network surrogate for the input-output behavior. The falsification is done using adversarial attack algorithms but is limited in its temporal operator support. Furthermore, it does not represent the dynamics as continuous time models. Online diffusion (OD) \cite{peltomaki2022wasserstein} uses online test generation, by denoising test vectors as parameterizations of piecewise constant input signals, based on generative adversarial networks. Moonlight \cite{nenzi2023moonlight} combines runtime monitoring with falsification and, in the ARCH-COMP 2024 competition, employs Bayesian optimization with trust regions to guide the search for violating inputs. EXAM-Net adopts an adversarial learning approach in which a discriminator network learns to predict robustness values from inputs, guiding the search toward likely violations.

Lately, more explicit formulations learn the system dynamics themselves, instead of an abstract robustness function mapping. This enables cheap simulations compared to the expensive SuT experiments. The surrogate dynamics can then be falsified, rather than the expensive SuT. Violating counterexamples are verified on the original SuT, making the algorithm sound with respect to the original system. ARIsTEO \cite{menghi2020approximation} estimates a set of models using system identification, and subjects them to black-box falsification. FreaK \cite{bak2024falsification} is another tool which uses Koopman operator theory and dynamic mode decomposition to estimate a linear function of observables. Although this does not provide an explicit dynamics model, it yields a linear surrogate model in a lifted observable space, enabling MILP solvers to compute the least robust trajectory. Lastly, FalConN uses neural ODEs to learn the differential dynamics \cite{kotz2026optimal}. The framework supports grey-box scenarios in which prior knowledge about dynamics is available or assumed. The falsification step is carried out by identifying a symbolic representation and solving a nonlinear optimal control problem to synthesize falsifying trajectories, given the specification. However, it is inherently single-modal and does not learn the discrete modes nor guard functions and transitions.

Finally, some data-driven tools focus on reachability or abstraction rather than optimization-based falsification. DryVR \cite{fan2017dryvr} learns probabilistic discrepancy functions to construct over-approximated reachability tubes. FalCAuN \cite{waga2020falsification} models the system under test as a black-box transition system and applies active automata learning to construct a Mealy machine abstraction, which is then analyzed using model checking techniques.

In conclusion, existing falsification frameworks either (i) treat the system as a black box and rely on gradient-free search, (ii) learn static input–output surrogates without modeling system dynamics, or (iii) approximate dynamics but do not expose differentiable structure for gradient-based optimization. Moreover, prior surrogate-based approaches primarily target purely continuous systems and do not address the challenges of learning and exploiting hybrid mode structure. This gap motivates the work presented in this paper.

\section{Preliminaries}
\subsection{Hybrid Automata}
We consider hybrid automata \cite{lygeros2003dynamical} as a tuple 
\begin{equation*}
    \mathcal{H} \triangleq \langle Q, X, F, \text{Inv}, R, E, G, \text{Init} \rangle
\end{equation*}
with finite set of modes $Q = \{q_1, \ldots, q_M\}$, $n-$dimensional continuous state space $X \subseteq \mathbb{R}^n$, transition edges $E \subseteq Q \times Q$, initial hybrid state $\text{Init} \subseteq X \times Q$, flow map $F = \{f_q: X \times U \times \mathbb{R}_{\geq 0} \rightarrow \mathbb{R}^n \mid q \in Q \}$, invariant $\text{Inv}: Q \rightarrow 2^X$, reset map $R: E \times X \to X$ and guard mapping $G: E\rightarrow 2^X$. 
The continuous state $x \in X$ evolves according to the vector field $f_q \in F$ conditioned on the discrete mode $q \in Q$ under invariant $\text{Inv}(q) \subseteq X$ and driven by a continuous input signal $u(t) \in U$
\begin{equation}
\begin{aligned}
    \dot{x} &= f_{q}(x, u, t), \quad x(t) \in \text{Inv}(q).  \\
\end{aligned}
\end{equation}

This state evolution is continuous per mode and piecewise continuous globally. Each transition edge $e \in E$ is a tuple $(q, g, r, q')$ consisting of the modes pre and post transition $q, q'$, along with the guard condition $g$ and reset map $r:$
\begin{equation}
    x\in G(e) \implies (q, x) \to (q',  R(e,x)).
\end{equation}

We denote a trajectory $\tau$ from $\mathcal{H}$ to be obtained by executing experiments on a hybrid system, either real-world or in simulation $\tau \gets \text{Exec}_{\mathcal{H}}(x_0, q_0, u(t))$.

\subsection{Signal Temporal Logic}
Signal temporal logic (STL) is a logic formalism which extends linear temporal logic with real-time and real-valued constraints for signals. Its grammar - given a set of atomic predicates $\mu$ - is recursively defined as 
\begin{equation*}
    \varphi ::= \mu \mid \neg \mu \mid \varphi \land \psi \mid \varphi \lor \psi \mid \square_{[a, b]}\psi \mid \lozenge_{[a, b]}\psi \mid \varphi \mathcal{U}_{[a, b]}\psi.
\end{equation*}
Here, $\square$, $\lozenge$ and $\mathcal{U}$ denote the globally, finally and until operators, respectively, and $[a, b]$ for $a, b \in \mathbb{R}_{0^+}, b\ge a$ define the closed interval time bounds.  

\begin{definition}[STL semantics]
The STL semantics of a signal $y$ at time $t$ are defined as
\begin{equation*}
    \begin{aligned}
        (y, t) & \vDash \mu \quad  & \iff &\mu(y(t)) > 0 \\
        (y, t) & \vDash \neg \mu \quad  &\iff &\neg((y, t) \vDash \mu) \\
        (y, t) & \vDash \varphi\land\psi & \iff & (y, t) \vDash \varphi \land (y, t) \vDash \psi \\
        (y, t) & \vDash \varphi \lor \psi & \iff & (y,t) \vDash \varphi \lor (y,t) \vDash \psi \\
        (y, t) & \vDash \square_{[a, b]} \varphi  & \iff & \forall t' \in [t + a, t+b], (y, t') \vDash \varphi \\
        (y, t) & \vDash \lozenge_{[a, b]} \varphi & \iff & \exists t' \in [t + a, t+b], (y, t') \vDash \varphi \\
        (y, t) & \vDash \varphi \mathcal{U}_{[a, b]} \psi & \iff & \exists t' \in [t+a, t+b], (y,t') \vDash \psi \\ & & & \land \forall t'' \in [t, t'], (y, t'') \vDash \varphi.
    \end{aligned}
\end{equation*}
\end{definition}
\begin{definition}[Robustness semantics]
For a scalar real-valued function $\rho:\mathbb{R}^{n_{signals} } \mapsto \mathbb{R}$ on a specification $\varphi$, the robustness value is a metric of how strongly $\varphi$ is satisfied or violated \cite{fainekos2009robustness}. The max semantics are given by
\begin{equation*}
    \begin{aligned}
        (y, t) \vDash \varphi & \iff \rho^\varphi(y, t) > 0 \\
        \rho^{\neg \varphi}(y, t) &= -\rho^\varphi(y, t) \\
        \rho^{\varphi_1 \land \varphi_2}(y, t) & = \min(\rho^{\varphi_1}(y, t), \rho^{\varphi_2}(y, t)) \\
        \rho^{\varphi_1 \lor \varphi_2}(y, t)& = \max(\rho^{\varphi_1}(y, t), \rho^{\varphi_2}(y, t)) \\
        \rho^{\lozenge_{[a, b]}\varphi}(y, t) &= \max_{t' \in [t+a, t+b]}(\rho^\varphi(y, t')) \\
        \rho^{\square_{[a, b]}\varphi}(y, t) &= \min_{t' \in [t+a, t+b]}(\rho^\varphi(y, t'))\\
        \rho^{\varphi_1\mathcal{U}_{[a, b]}\varphi_2}(y, t) &= \max_{t' \in [t+a, t+b]}(\min([\rho^{\varphi_2}(y, t'), \\ &\min_{t'' \in [t+a, t']}(\rho^{\varphi_1}(y, t''))])).
    \end{aligned}
\end{equation*}
\end{definition}

%% file: Sections/Method.tex
\section{Falsification using NHAs and Optimal Control}
We expand the work in \cite{kotz2026optimal} – which previously assumed purely continuous dynamics – to incorporate hybrid automata by employing Neural Hybrid Automata (NHA) for learning \cite{poli2021neural}. This approach requires the user to make an assumption of the number of modes $M$ in the system, although empirical results suggest that redundant modes collapse into one. Furthermore, approximate transition times or heuristics to determine them from trajectories are needed. In ideal cases, an observer can indicate when transitions occur, but in general we rely on change-point detection methods. Source and target mode information of transitions is not required. 

The learned NHA is then subjected to a gradient-based optimal control formulation, in which robustness is smoothly approximated and thus fully differentiable with respect to input signals. If the objective value of the optimization problem is below zero, the surrogate is falsified and the counterexample is subsequently verified on the original SuT to avoid spurious results.

The algorithm is not complete, but remains sound, as candidate counterexamples are always evaluated on the original SuT.


\subsection{Neural Hybrid Automata learning}
The NHA consists of a mode encoder $\varepsilon$ parameterized by $\omega$, trained to separate trajectory segments, and a piecewise-continuous flow model. For a segment $\tau_k$ active over $[t_k, t_{k+1}]$, the dynamics are represented as a neural ODE \cite{chen2018neural} with potential a priori knowledge of the system dynamics:
\begin{equation}
    \dot{x} = f_{\theta_{q_k}}(x, u, t) + f^{\mathrm{known}}_{q_k}(x, u, t), \quad t \in [t_k, t_{k+1}],
\end{equation}
where $q_k$ is the discrete mode for the segment. The unknown dynamics $f_{\theta_{q_k}}$ are parameterized by a universal approximator that is Lipschitz in $x$ and continuous in $t$, such as an MLP with tanh activations, guaranteeing existence and uniqueness of solutions by the Picard-Lindelöf theorem.

During training, we utilize a straight-through estimator, executing a hard mode selection during the forward pass, while gradients are computed using a differentiable sum of vector fields weighted by the softmax-derived mode probabilities $p^k$:
\begin{equation}
    \begin{aligned}
        p^{k} &= \text{softmax(}\varepsilon(\tau_k, \omega)),  \\
        \hat{q} &= \arg \max_{q\in Q} p^k_q, \\
        \dot{x} &=  f_{\theta_{\hat{q}}}(x, u, t) + f^{\mathrm{known}}_{\hat{q}}(x, u, t), \\
        \frac{\partial \dot{x}}{\partial \theta}
        &\approx
        \sum_{q \in Q} p^k_q
        \frac{\partial}{\partial \theta}
        \left[
            f_{\theta_q}(x,u,t) + f^{\mathrm{known}}_q(x,u,t)
        \right],
    \end{aligned}
\end{equation}
where $p_q^k$ is the estimated probability of segment $k$ being attributed to latent mode $q$ such that
\begin{equation}
p^k \in \Delta^Q, \quad
\Delta^Q = \left \{p \in \mathbb{R}^Q \mid p_q \ge 0, \sum_{q \in Q} p_q = 1 \right \}.
\end{equation}

We train the encoder and neural ODEs jointly to cluster segments into latent modes. Neural ODE gradients can be computed by solving the adjoint sensitivity equation \cite{pontryagin2018mathematical} backwards in time after having computed the loss at observation points.

To mitigate mode collapse in segment boundaries, we add a mode separation penalty term to the total loss, defined as the inner product of the mode probability distributions
\begin{equation}\label{eq:loss}
    \mathcal{L}^\tau = \sum_{k=1}^K \mathcal{L}^{(\tau_k)}_{\text{rec}} + \lambda_{\text{sep}} \sum_{k=1}^{K-1} \langle p^k, p^{k+1} \rangle ,
\end{equation}
where $\lambda_{\text{sep}}$ is a regularization factor responsible for separating subsequent segments and discouraging mode collapse, whereas $\mathcal{L}_{\text{rec}}$ is the MSE reconstruction error of subtrajectory $\tau_k$. 

The guard functions are learned using a hierarchical model selection strategy, starting from a linear inductive structure and only resorting to nonlinearity if the linear hypothesis proves insufficient. Determining linear guards for each observed transition corresponds to solving the hyperplane $a_1x_1 + \ldots + a_nx_n + b = 0$ that defines the state $x$ at the respective transition. The parameters $w = (a_1, \ldots, a_n, b)$ are estimated as a solution to $w = \text{null}([X, 1])$. Determining an $n-$dimensional hyperplane requires making $n$ linearly independent observations per unique transition.

\subsection{Gradient-based optimal control}
With a hybrid surrogate dynamical system $f_\mathcal{H}$ available, we formulate the falsification problem as an optimal control problem, minimizing the robustness. The robustness objective gradient with respect to the input signal is obtained by differentiating through the surrogate simulator which can be done efficiently using automatic differentiation. 

The robustness semantics of STL are in general non-differentiable due to $\max$- and $\min$-operators. However, because the robustness gradient $\nabla_{u} \rho^{\neg \varphi}$ is required, the required operators are smoothly approximated using a log-sum exponential as introduced in \cite{pant2017smooth}
\begin{equation}
    \begin{aligned}
        \max ([a_1, \ldots, a_m]) & \approx \frac{1}{s} \log \left  (\sum_{i=1}^m e^{sa_i} \right )\\
        \min ([a_1, \ldots, a_m]) & \approx -\frac{1}{s} \log \left ( \sum_{i=1}^m e^{-sa_i} \right ),
    \end{aligned}
\end{equation}
where $s$ is a smoothing parameter. As $s \rightarrow \infty$, the operators become exact, offering asymptotic completeness of the robustness semantics.

A least-robust input signal is acquired by solving the optimal control formulation subject to the system dynamics and input constraints:

\begin{equation}
\begin{aligned}
    & \min_{u(\cdot)} \: \rho^{ \varphi}(x(\cdot), u(\cdot)) \\
    &\text{s.t. } \dot{x} = f_\mathcal{H}(x, u, t) \\
    u_{\text{min}} & \leq u(t) \leq u_{\text{max}}, \quad \forall t \in [0, T].\\
\end{aligned}
\end{equation}
This can be done in a multitude of ways, but given that the objective function is nonconvex w.r.t. inputs $u$, we employ a single-shooting quasi-Newton formulation to search for a falsifying trajectory using the L-BFGS algorithm. This enables approximate Hessian information, improving convergence near optima. The resulting optimal control problem induced by the smoothed STL specification is highly non-convex and therefore sensitive to initialization. We adopt the initialization strategy of \cite{kotz2026optimal}, in which a batch of input parameters is sampled randomly and evaluated on the surrogate model. Each sample is scored using its robustness value normalized by its Euclidean distance to the training input parameters $\phi_{\mathrm{data}}$. The initialization is then selected as
\begin{equation}
    \phi_o = \arg \min_{\phi \in \{\phi_1, \ldots, \phi_{P}\}} 
    \frac{\rho^\varphi (\mathrm{Exec}_{\mathcal{H}}(x_0, q_0, u_\phi(t)))}{\| \phi - \phi_{\mathrm{data}} \|_2}.
\end{equation}
This heuristic favors input signals that are both likely to violate the specification on the surrogate and sufficiently distinct from the training data, improving exploration of the optimization landscape.

%% file: Sections/Evaluation.tex
\section{Evaluation Setup}
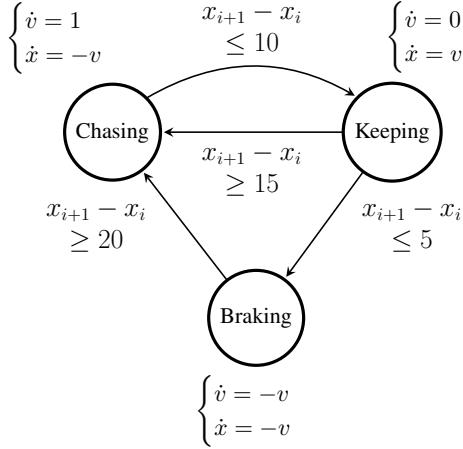
\begin{figure}[H]
    \centering
        \centering
        \input{Figures/automaton} 
        \caption{Hybrid automaton for cruise policy including guard conditions as functions of relative position $x_{i+1} - x_i$. The initial state is keeping.}
        \label{fig:automaton}
\end{figure}

We evaluate our method on the chasing cars system in the ARCH-COMP 2024 falsification benchmark \cite{khandait2024arch}. The chasing cars system \cite{hu2000towards} is set up with five cars driving along a straight road. The input signals are the throttle and brake to the leading car, which is described by a second order differential equation. The remaining four cars are described by a 3-mode hybrid automaton with the ``keeping``, ``chasing`` and ``braking`` modes used to keep distance while not lagging behind, as shown in Figure \ref{fig:automaton}. The associated STL-specifications for the system are provided in Table \ref{tab:placeholder} and specify both collision-free and proximity behaviors. We use the same benchmark-model as the tools in \cite{khandait2024arch}, which differs in some modeling details from the original formulation in \cite{hu2000towards}.\footnote{Evaluations on a corrected version were also made, which resulted in non-challenging instances in which all specifications are almost immediately falsified.}

\begin{table}[h!]
    \caption{The STL safety specifications for chasing cars constitutes a mix of collision-avoidance and proximity specifications between the positions of chasing car $y_{i+1}$ and leading car $y_i$.}
    \centering
    \input{Benchmarks/specifications} 
\end{table}

The competition benchmarks feature two different instances; piecewise-constant and arbitrary continuous parameterization of the inputs. To ensure consistent comparison with other tools, we evaluate our method on the fixed parameterization of the inputs, with with switching interval of $5$ seconds. 

We initialize our data via uniform-random sampling of the segment magnitudes. For model structure, we assume NHAs with 3 latent modes, which is more than those reachable in the underlying system, whereas the leading car is parameterized by a single mode. To limit training overhead and avoid converging to local optima, we run the experiments with a memory of the last five SuT trajectories, discarding prior ones. Lastly, the smoothing coefficient of the robustness semantics is set to $s=2$, balancing smoothness and exactness.

We evaluate our method over 5 repeated runs to mitigate the influence of randomness. We set a significantly lower simulation budget (20) than the benchmarks (1500) to limit runtime. 

%% file: Benchmarks/specifications.tex
    \begin{tabular}{c|c} \hline
        Spec. & STL \\ \hline
        CC1 & $\square_{[0, 100]} y_5 - y_4 \leq 40$ \\ \hline
        CC2 & $\square_{[0, 70]}\lozenge_{[0, 30]} y_5 - y_4 \geq 15$ \\ \hline
        CC3 & \begin{tabular}{c}
             $\square_{[0, 80]} ((\square_{[0, 20]} y_2 - y_1 \leq 20)$ \\
             $\lor (\lozenge_{[0, 20]} y_5 - y_4 \geq 40))$
        \end{tabular}  \\ \hline
        CC4 & $\square_{[0, 65]} \lozenge_{[0, 30]}\square_{[0, 5]} y_5 - y_4 \geq 8$ \\ \hline
        CC5 & \begin{tabular}{c}
              $\square_{[0, 72]}\lozenge_{[0, 8]} ((\square_{[0, 5]}y_2-y_1 \geq 9)$\\
              $\rightarrow (\square_{[5, 20]}y_5 - y_4 \geq 9))$
        \end{tabular}   \\ \hline
        CCx & $\bigwedge_{i=1..4} \square_{[0, 50]}(y_{i+1} - y_i >7.5 )$ \\ \hline
    \end{tabular}
    \label{tab:placeholder}

%% file: Figures/automaton.tex
\definecolor{green}{RGB}{40,188,0}
\definecolor{red}{RGB}{155,0,0}
\definecolor{darkgreen}{RGB}{8,137,0}
\definecolor{darkblue}{RGB}{9,52,226}
\definecolor{state3background}{RGB}{180, 185, 193}

\usetikzlibrary{arrows.meta,automata,calc,positioning, shapes.geometric, calc}

\tikzset{
    state1/.style={
           circle,
           draw=black, very thick,
           minimum height=2em,
           inner sep=2pt,
           minimum size=2em,
           font=\huge
           },
    state2/.style={
           circle,
           draw=black, very thick,
           minimum height=2em,
           inner sep=2pt,
           minimum size=2em,
           font=\huge
           },
    state3/.style={
           circle,
           draw=black, very thick,
           minimum height=2em,
           inner sep=2pt,
           minimum size=2em,
           font=\huge
           }
}

\begin{tikzpicture}[->,>=stealth,shorten >=1pt,auto,semithick, transform shape, scale=0.6]

\node[state1] (1) [scale=0.7]{
    \begin{tabular}{c}
        Chasing
    \end{tabular}};

\node[state2] (2) [right=4.0cm of 1, scale=0.7] {
    \begin{tabular}{c}
        Keeping
    \end{tabular}
    };

\coordinate (tmp) at ($(2.south) - (0,3cm) - (3cm,0)$);

\node[state3] (3) at (tmp) [scale=0.7] {
    \begin{tabular}{c}
        Braking \\
    \end{tabular}
    };

\coordinate (m1_eq) at ($(1.north) - (1cm, 0) + (0, 1cm)$);
\coordinate (m2_eq) at ($(2.north) + (1cm, 1cm)$);
\coordinate (m3_eq) at ($(3.south) - (0, 1cm)$);

\node at (m1_eq) [scale=1.5] {$\begin{cases}\dot{v} = 1 \\ \dot{x} = -v\end{cases}$};
\node at (m2_eq) [scale=1.5] {$\begin{cases}\dot{v} = 0 \\ \dot{x} = v \end{cases}$};
\node at (m3_eq) [scale=1.5] {$\begin{cases} \dot{v} = -v\\ \dot{x} = -v \end{cases}$};

\path

    (2) edge [below] node[scale=1.0, font=\LARGE] {
        \begin{tabular}{c}
        $x_{i+1} - x_i$ \\ $\geq 15$ 
        \end{tabular}
        } (1)
    (2) edge [right] node[scale=1.0, xshift=0.5cm, font=\LARGE] {
        \begin{tabular}{c}
            $x_{i+1} - x_i$ \\
            $\leq 5$
        \end{tabular}
        } (3)
    (3) edge [left] node[scale=1.0, xshift=-0.5cm, font=\LARGE] {
        \begin{tabular}{c}
            $x_{i+1} - x_i$ \\ $\geq 20$ 
        \end{tabular}
    } (1)
    (1.north east) edge[bend left] node[midway, above, scale=1.0, font=\LARGE]
    {
      \begin{tabular}{c}
        $x_{i+1} - x_{i}$ \\
        $\le 10$
      \end{tabular}
    } (2.north west);

\end{tikzpicture}

%% file: Sections/Results.tex
\section{Results}

We evaluate the proposed NHA-based falsification method on the ARCH-COMP 2024 chasing cars benchmark, comparing against all participating tools in Instance 2 (Table~\ref{tab:combined_transposed}). Performance is measured in terms of the number of successful falsifications (FR) and the number of SuT executions required to find a counterexample (mean $\bar{S}$ and median $\tilde{S}$).

Across specifications, the proposed method consistently achieves successful falsification within the available simulation budget in cases where a counterexample is found. In CC$_1$, CC$_2$, CC$_3$, and CC$_5$, our method attains comparable or lower numbers of SuT evaluations than several baseline methods, despite using a significantly smaller evaluation budget (20 SuT executions per run versus up to 1500 in the benchmark setting).

In particular, for CC$_5$, our method achieves the lowest median and mean number of SuT executions among all evaluated tools, indicating a high sample efficiency. Similarly, for CC$_1$, CC$_2$ and CC$_3$, the number of required simulations is competitive with or lower than most gradient-free and surrogate-based approaches, while maintaining consistent falsification success across repeated runs.

For CC$_4$ and CC$_x$, the limited simulation budget leads to incomplete falsification within the allocated evaluation limit. In these cases, several baseline methods achieve higher success rates or require more evaluations to identify counterexamples, indicating that these specifications generally require larger exploration budgets across methods.

\begin{figure}[h!]
    \centering
    \includegraphics[width=1.0\linewidth]{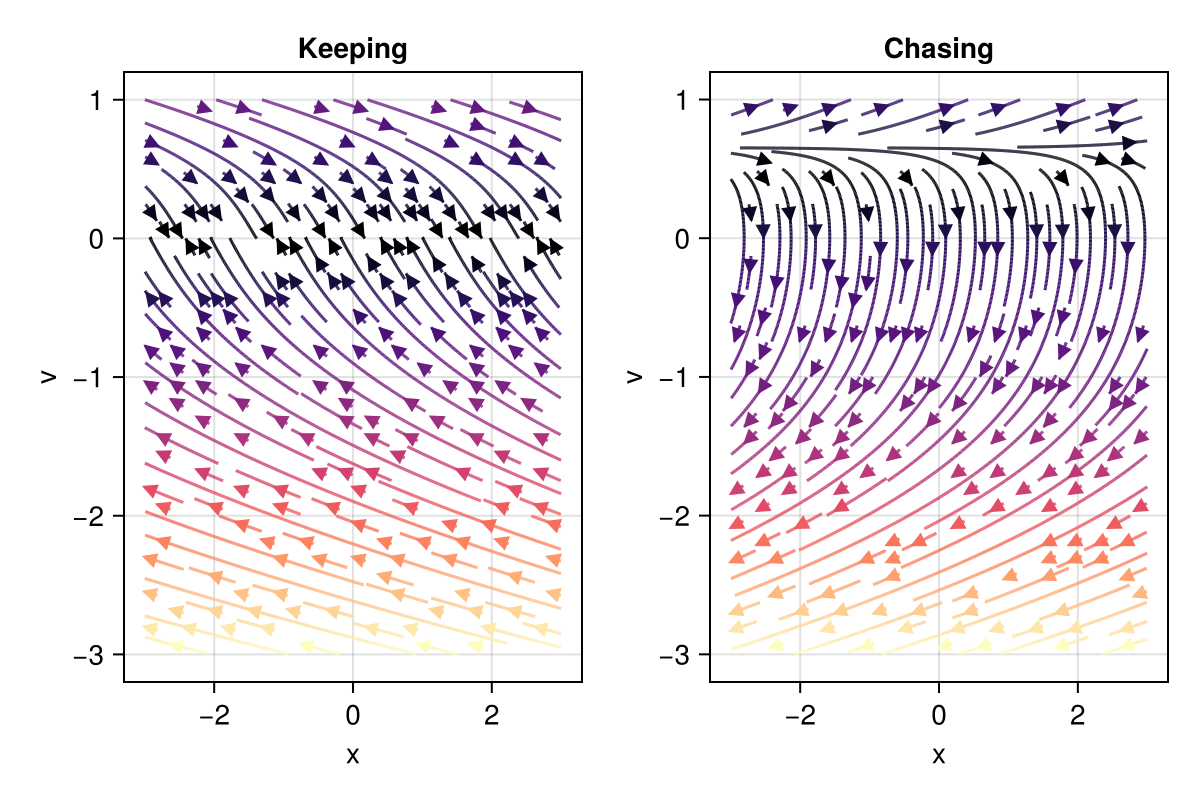}
    \caption{The learned NHA vector fields used to successfully falsify the SuT. Note that the unreachable braking mode has been omitted.}
    \label{fig:placeholder}
\end{figure}

At inspection of the learned NHAs vector fields (Figure \ref{fig:placeholder}), we find that they successfully prune the redundant and unreachable mode, indicating a robustness to over-approximation. This is similar to the findings in \cite{poli2021neural} and suggests the method is insensitive to overapproximation of the number of modes assumed a priori. 

Overall, the results indicate that gradient-based optimization on learned hybrid surrogates can reduce the number of required SuT evaluations in several benchmark instances, while remaining competitive with state-of-the-art falsification tools under a significantly tighter evaluation budget.

\begin{table}[t!]
    \centering
    \input{Benchmarks/instance2}
\end{table}

%% file: Benchmarks/instance2.tex
\caption{Performance of our approach in the modified chasing cars model vs the participating tools in the ARCH-COMP 2024 competition, instance 2. We measure the number of successful falsifications FR, mean $\bar{S}$, and median $\tilde{S}$ number of SuT executions to find a counterexample, and evaluate our tool with a simulation budget of 20, contrary to the benchmark evaluation.}
\resizebox{\columnwidth}{!}{
\begin{tabular}{c|ccc|ccc}
\textbf{Tool} &
FR & $\bar{S}$ & $\tilde{S}$ &
FR & $\bar{S}$ & $\tilde{S}$ \\
\hline
& \multicolumn{3}{c}{\textbf{CC$_1$}} & \multicolumn{3}{c}{\textbf{CC$_4$}} \\
\hline
Our Approach   & 5/5 & 5.6 & 5.0 & 0/5 & - & - \\
UR             & 10/10 & 16.4 & 9.5 & 0/10 & - & - \\
ARIsTEO        & 10/10 & 11.3 & 11.5 & 0/10 & - & - \\
ATheNA         & 10/10 & 60.5 & 30.0 & 2/10 & 514.0 & 514.0 \\
EXAM-Net       & 10/10 & 36.1 & 36.0 & 0/10 & - & - \\
FORESEE        & 10/10 & 21.8 & 21.5 & 7/10 & 680.6 & 732.0 \\
FReaK          & 10/10 & \textbf{3.0} & \textbf{2.0} & 10/10 & \textbf{1349.9} & \textbf{1201.0} \\
NNFal          & 9/10 & 1.22 & 1 &  &  &  \\
OD             & 10/10 & 61.2 & 53.0 & 0/10 & - & - \\
$\psi$--TaLiRo & 10/10 & 18.0 & 10.5 & 1/10 & 1387.3 & 1500.0 \\
\hline
& \multicolumn{3}{c}{\textbf{CC$_2$}} & \multicolumn{3}{c}{\textbf{CC$_5$}} \\
\hline
Our Approach   & 5/5 & 6.6 & 4.0 & 5/5 & \textbf{10.6} & \textbf{6.0} \\
UR             & 10/10 & 12.4 & 13.0 & 10/10 & 37.4 & 22.0 \\
ARIsTEO        & 10/10 & 10.5 & 10.0 & 10/10 & 29.1 & 22.5  \\
ATheNA         & 10/10 & 94.4 & 106.5 & 9/10 & 112.0 & 91.0 \\
EXAM-Net       & 9/10 & 398.6 & 364 & 10/10 & 78.2 & 106 \\
FORESEE        & 8/10 & 224.3 & 19.5 &  10/10 & 88.7 & 25.5 \\
FReaK          & 10/10 & \textbf{3.3} & \textbf{3.0} &  10/10 & 47.5 & 34.0 \\
NNFal          &  &  &  &  &  &  \\
OD             & 10/10 & 109.3 & 44.0 & 10/10 & 55.7 & 53.0 \\
$\psi$--TaLiRo & 10/10 & 14.8 & 9.0 & 10/10 & 32.2 & 36.5 \\
\hline
& \multicolumn{3}{c}{\textbf{CC$_3$}} & \multicolumn{3}{c}{\textbf{CC$_x$}} \\
\hline
Our Approach   & 5/5 & 3.4 & \textbf{2.0} & 1/5 & 19.0 & 19.0 \\
UR             & 10/10 & 19.6 & 21.0 & 6/10 & 396.7 & 284.5 \\
ARIsTEO        & 10/10 & 18.8 & 12.5 & 9/10 & 610.4 & 465.0 \\
ATheNA         & 10/10 & 119.2 & 95.5 & 5/10 & 86.4 & 84.0 \\
EXAM-Net       & 10/10 & 14.9 & 10/10 & 10/10 & 448.1 & 390.5 \\
FORESEE        & 10/10 & 36.1 & 11.5 & 10/10 & \textbf{228.0} & \textbf{189.5} \\
FReaK          & 10/10 & \textbf{2.6} & \textbf{2.0} & 7/10 & 1723.6 & 938.0 \\
NNFal          &  &  &  &  &  &  \\
OD             & 10/10 & 35.2 & 31.0 & 2/10 & 261.5 & 261.0 \\
$\psi$--TaLiRo & 10/10 & 11.4 & 10.5 & 10/10 & 244.2 & 225.0 \\
\hline
\end{tabular}}
\label{tab:combined_transposed}

%% file: Sections/Conclusions.tex
\section{Conclusions}
In this work, we have extended the FalConN surrogate-based falsification method  \cite{kotz2026optimal} to incorporate hybrid systems by incorporating NHAs with gradient-based optimal control. The approach relies on learning a differentiable piecewise-continuous automaton with a vectorfield per latent mode. Within fixed mode-sequences, the dynamics remain locally differentiable, enabling gradient-based optimal control, using smoothed STL robustness.

Experimental results on the ARCH-COMP 2024 chasing cars benchmark demonstrate that our method not only correctly discovers a reduced effective automaton structure to the chasing cars model, but also achieves competitive or improved sample-efficiency relative to existing tools on many of the specifications, under a significantly reduced simulation budget. 

Future work includes extending the method to more expressive trajectory optimization schemes, such as multiple-shooting methods for each branch of transitions as well as integrating hybrid gradient methods to search more efficiently for least-robust inputs.

%% file: root.bib
@inproceedings{bak2024falsification,
  title={Falsification using reachability of surrogate koopman models},
  author={Bak, Stanley and Bogomolov, Sergiy and Hekal, Abdelrahman and Kochdumper, Niklas and Lew, Ethan and Mata, Andrew and Rahmati, Amir},
  booktitle={Proceedings of the 27th ACM International Conference on Hybrid Systems: Computation and Control},
  pages={1--13},
  year={2024}
}

@article{ramezani2025falsification,
  title={Falsification of cyber-physical systems using Bayesian optimization},
  author={Ramezani, Zahra and {\v{S}}ehi{\'c}, Kenan and Nardi, Luigi and {\AA}kesson, Knut},
  journal={ACM Transactions on Embedded Computing Systems},
  volume={24},
  number={3},
  pages={1--23},
  year={2025},
  publisher={ACM New York, NY}
}

@article{chen2018neural,
  title={Neural ordinary differential equations},
  author       = {Tian Qi Chen and
                  Yulia Rubanova and
                  Jesse Bettencourt and
                  David Duvenaud},
  journal={Advances in neural information processing systems},
  volume={31},
  year={2018}
}

@book{pontryagin2018mathematical,
  title={Mathematical theory of optimal processes},
  author={Pontryagin, Lev Semenovich},
  year={2018},
  publisher={Routledge}
}

@inproceedings{menghi2020approximation,
  title={Approximation-refinement testing of compute-intensive cyber-physical models: An approach based on system identification},
  author={Menghi, Claudio and Nejati, Shiva and Briand, Lionel and Parache, Yago Isasi},
  booktitle={Proceedings of the ACM/IEEE 42nd International Conference on Software Engineering},
  pages={372--384},
  year={2020}
}

@inproceedings{khandait2024arch,
  title={Arch-comp 2024 category report: Falsification},
  author={Khandait, Tanmay and Formica, Federico and Arcaini, Paolo and Chotaliya, Surdeep and Fainekos, Georgios and Hekal, Abdelrahman and Kundu, Atanu and Lew, Ethan and Loreti, Michele and Menghi, Claudio and others},
  booktitle={International Workshop on Applied Verification for Continuous and Hybrid Systems},
  pages={122--144},
  year={2024},
  organization={EasyChair}
}

@inproceedings{donze2010breach,
  title={Breach, a toolbox for verification and parameter synthesis of hybrid systems},
  author={Donz{\'e}, Alexandre},
  booktitle={International Conference on Computer Aided Verification},
  pages={167--170},
  year={2010},
  organization={Springer}
}

@inproceedings{kundu2024data,
  title={Data-driven falsification of cyber-physical systems},
  author={Kundu, Atanu and Gon, Sauvik and Ray, Rajarshi},
  booktitle={Proceedings of the 17th Innovations in Software Engineering Conference},
  pages={1--5},
  year={2024}
}

@inproceedings{pant2017smooth,
  title={Smooth operator: Control using the smooth robustness of temporal logic},
  author={Pant, Yash Vardhan and Abbas, Houssam and Mangharam, Rahul},
  booktitle={2017 IEEE Conference on Control Technology and Applications (CCTA)},
  pages={1235--1240},
  year={2017},
  organization={IEEE}
}

@inproceedings{annpureddy2011s,
  title={S-taliro: A tool for temporal logic falsification for hybrid systems},
  author={Annpureddy, Yashwanth and Liu, Che and Fainekos, Georgios and Sankaranarayanan, Sriram},
  booktitle={International Conference on Tools and Algorithms for the Construction and Analysis of Systems},
  pages={254--257},
  year={2011},
  organization={Springer}
}

@inproceedings{peltomaki2022wasserstein,
  title={Wasserstein generative adversarial networks for online test generation for cyber physical systems},
  author={Peltom{\"a}ki, Jarkko and Spencer, Frankie and Porres, Ivan},
  booktitle={Proceedings of the 15th Workshop on Search-Based Software Testing},
  pages={1--5},
  year={2022}
}

@article{nenzi2023moonlight,
  title={MoonLight: a lightweight tool for monitoring spatio-temporal properties},
  author={Nenzi, Laura and Bartocci, Ezio and Bortolussi, Luca and Silvetti, Simone and Loreti, Michele},
  journal={International Journal on Software Tools for Technology Transfer},
  volume={25},
  number={4},
  pages={503--517},
  year={2023},
  publisher={Springer}
}

@inproceedings{zhang2021effective,
  title={Effective hybrid system falsification using Monte Carlo tree search guided by {QB}-robustness},
  author={Zhang, Zhenya and Lyu, Deyun and Arcaini, Paolo and Ma, Lei and Hasuo, Ichiro and Zhao, Jianjun},
  booktitle={International Conference on Computer Aided Verification},
  pages={595--618},
  year={2021},
  organization={Springer}
}

@inproceedings{waga2020falsification,
  title={Falsification of cyber-physical systems with robustness-guided black-box checking},
  author={Waga, Masaki},
  booktitle={Proceedings of the 23rd International Conference on Hybrid Systems: Computation and Control},
  pages={1--13},
  year={2020}
}

@article{formica2023search,
  title={Search-based software testing driven by automatically generated and manually defined fitness functions},
  author={Formica, Federico and Fan, Tony and Menghi, Claudio},
  journal={ACM Transactions on Software Engineering and Methodology},
  volume={33},
  number={2},
  pages={1--37},
  year={2023},
  publisher={ACM New York, NY}
}

@article{poli2021neural,
  title={Neural hybrid automata: Learning dynamics with multiple modes and stochastic transitions},
  author={Poli, Michael and Massaroli, Stefano and Scimeca, Luca and Chun, Sanghyuk and Oh, Seong Joon and Yamashita, Atsushi and Asama, Hajime and Park, Jinkyoo and Garg, Animesh},
  journal={Advances in Neural Information Processing Systems},
  volume={34},
  pages={9977--9989},
  year={2021}
}

@inproceedings{fan2017dryvr,
  title={DryVR: Data-driven verification and compositional reasoning for automotive systems},
  author={Fan, Chuchu and Qi, Bolun and Mitra, Sayan and Viswanathan, Mahesh},
  booktitle={International Conference on Computer Aided Verification},
  pages={441--461},
  year={2017},
  organization={Springer}
}

@inproceedings{hu2000towards,
  title={Towards a theory of stochastic hybrid systems},
  author={Hu, Jianghai and Lygeros, John and Sastry, Shankar},
  booktitle={International Workshop on Hybrid Systems: Computation and Control},
  pages={160--173},
  year={2000},
  organization={Springer}
}

@article{fainekos2009robustness,
  title={Robustness of temporal logic specifications for continuous-time signals},
  author={Fainekos, Georgios E and Pappas, George J},
  journal={Theoretical Computer Science},
  volume={410},
  number={42},
  pages={4262--4291},
  year={2009},
  publisher={Elsevier}
}

@misc{kotz2026optimal,
  title={Optimal Control-Based Falsification of Learnt Dynamics via Neural ODEs and Symbolic Regression},
  author={K{\"o}tz, Lasse and Sj{\"o}berg, Jonas and {\AA}kesson, Knut},
  booktitle = {Proceedings of the 29th ACM International Conference on Hybrid Systems: Computation and Control},
  year      = {in press},
  note      = {accepted for publication, to appear in HSCC 2026 proceedings},
  doi       = {10.1145/3801146.3805672}, 
  address   = {Saint Malo, France},
  month     = {May}
}

@article{lygeros2003dynamical,
  title={Dynamical properties of hybrid automata},
  author={Lygeros, John and Johansson, Karl Henrik and Simic, Slobodan N and Zhang, Jun and Sastry, S Shankar},
  journal={IEEE Transactions on automatic control},
  volume={48},
  number={1},
  pages={2--17},
  year={2003},
  publisher={IEEE}
}
